\pdfoutput=1
\documentclass[final,twocolumn,prl,aps,floats,amsfonts,amssymb,showpacs,superscriptaddress]{revtex4-1}
\usepackage{graphicx}
\usepackage{color}
\usepackage{gensymb}
\usepackage{enumerate}
\usepackage{stmaryrd}
\usepackage{amssymb}
\usepackage{amsmath}
\usepackage{wasysym}
\usepackage{hyperref}

\begin{document}

\title{Experimental evidence of a phase transition in a closed turbulent flow}

\author{P.-P. Cortet}
\altaffiliation[Present address:~]{Laboratoire FAST, CNRS UMR
7608, Universit\'e Paris-Sud, 91405 Orsay, France}
\author{A. Chiffaudel}
\author{F. Daviaud}
\author{B. Dubrulle}
\affiliation{CEA, IRAMIS, SPEC, CNRS URA 2464, Groupe
Instabilit\'{e}s \& Turbulence, 91191 Gif-sur-Yvette, France}

\date{\today}


\begin{abstract}

We experimentally study the susceptibility to symmetry breaking of
a closed turbulent von K\'{a}rm\'{a}n swirling flow from $Re =
150$ to $Re \simeq 10^{6}$. We report a divergence of this
susceptibility at an intermediate Reynolds number $Re = Re_\chi
\simeq 90\,000$ which gives experimental evidence that such a
highly space and time fluctuating system can undergo a ``phase
transition''. This transition is furthermore associated with a
peak in the amplitude of fluctuations of the instantaneous flow
symmetry corresponding to intermittencies between spontaneously
symmetry breaking metastable states.
\end{abstract}

\maketitle

Phase transitions are ubiquitous in physical systems and generally
are associated to symmetry breakings. For example, ferromagnetic
systems are well known to undergo a phase transition from
paramagnetism to ferromagnetism at the Curie temperature $T_c$.
This transition is associated with a symmetry breaking from the
disordered paramagnetic ---associated to a zero magnetization---
toward the ordered ferromagnetic phase
---associated to a finite magnetization--- \cite{landau}.
In the vicinity of $T_c$, a singular behaviour
characterized by critical exponents is observed, \emph{e.g.} for
the magnetic susceptibility to an external magnetic field. In the
context of fluid dynamics, symmetry breaking also governs the
transition to turbulence, that usually proceeds, as the Reynolds
number $Re$ increases, through a sequence of bifurcations breaking
successively the various symmetries allowed by the Navier-Stokes
equations coupled to the boundary conditions \cite{manneville}.
Finally, at large Reynolds number, when the fully developed
turbulent regime is reached, it is commonly admitted that all the
broken symmetries are restored in a statistical sense, the
statistical properties of the flow not depending anymore on $Re$
\cite{frisch}. However, recent experimental studies of turbulent
flows have disturbed this vision raising intriguing features such
as finite lifetime turbulence \cite{hof} ---questioning the
stability of the turbulent regime--- and possible existence of
turbulent transitions
\cite{castaing,mujica,lohse2009,tabeling96,tabeling02,ravelet2004,ravelet2008}.
Consequently, despite turbulent flows are intrinsically
out-of-equilibrium systems, one may wonder whether the observed
transitions can be interpreted in terms of phase transitions with
a symmetry-breaking or susceptibility divergence signature. In
this paper, we introduce a susceptibility to symmetry breaking in
a von K\'{a}rm\'{a}n turbulent flow and investigate its evolution
as $Re$ increases from $150$ to $10^6$ using stereoscopic Particle
Image Velocimetry (PIV). We observe a divergence of susceptibility
at a critical Reynolds number $Re = Re_\chi \simeq 90\,000$ which
sets the threshold for a possible turbulent ``phase transition''.
Moreover, this divergence is associated with a peak in the
amplitude of the fluctuations of the flow instantaneous symmetry.

\begin{figure}
\includegraphics[width=0.6\columnwidth]{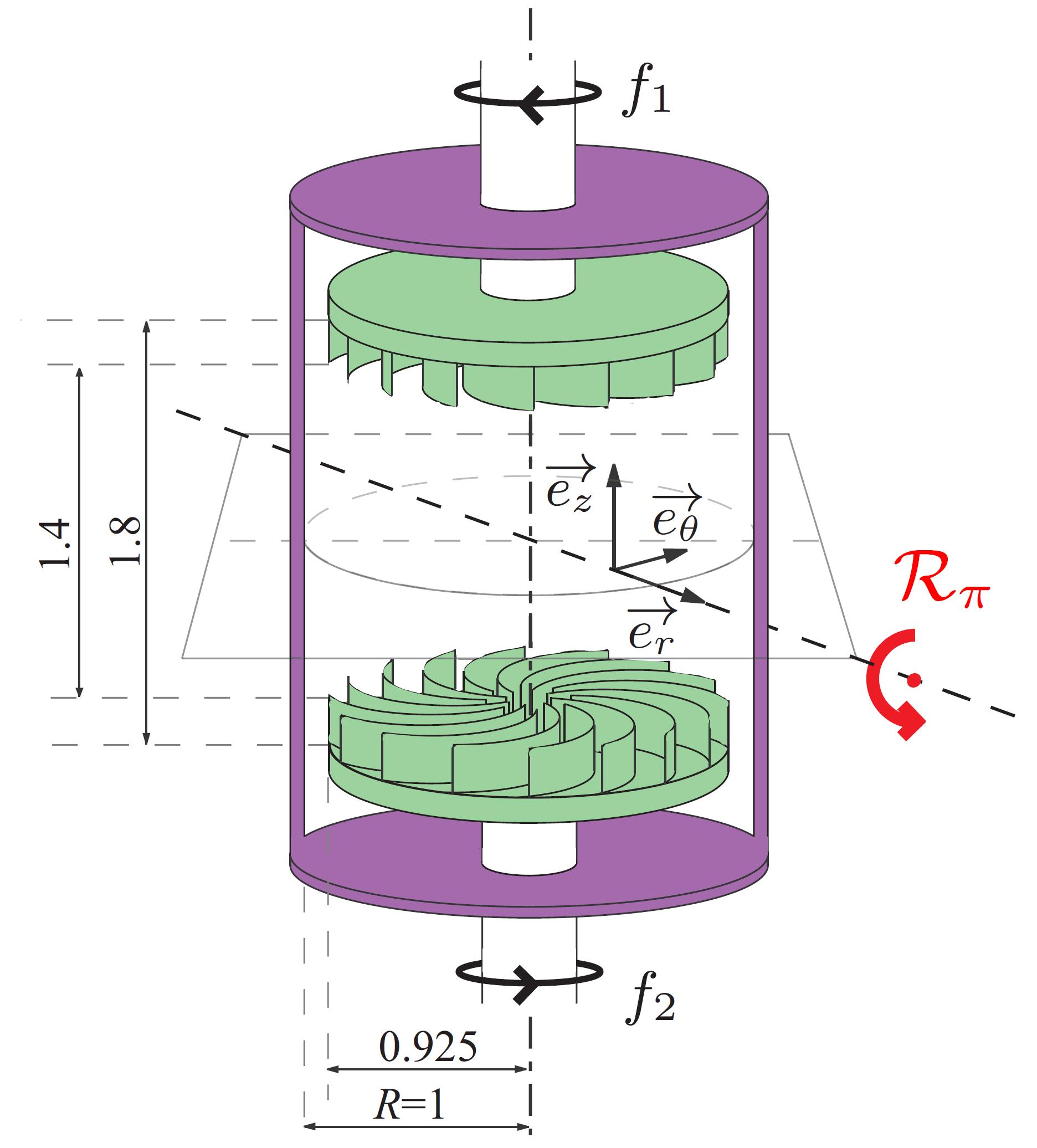}
\caption{Schematic view of the experimental setup and the
impellers blade profile. The arrow on the shaft indicates the
impeller rotation sense studied. Symmetry~: the system is
symmetric regarding any $\cal{R}_\pi$-rotation of angle $\pi$
around any line of the equatorial plane which crosses the rotation
axis.}\label{fig1}
\end{figure}

Our experimental setup consists of a Plexiglas cylinder of radius
$R=100$~mm filled up with either water or water-glycerol mixtures.
The fluid is mechanically stirred by a pair of coaxial impellers
rotating in opposite sense (Fig. \ref{fig1}). The impellers are
flat disks of radius $0.925\,R$, fitted with 16 radial blades of
height $0.2\,R$ and curvature radius $0.4625 \,R$. The disks inner
surfaces are $1.8\,R$ apart setting the axial distance between
impellers from blades to blades to $1.4\,R$. The impellers rotate,
with the convex face of the blades pushing the fluid forward,
driven by two independent brushless 1.8~kW motors. The rotation
frequencies $f_1$ and $f_2$ can be varied independently from $1$
to $12$~Hz. Velocity measurements are performed with a
stereoscopic PIV system provided by DANTEC Dynamics. The data
provide the radial $u_r$, axial $u_z$ and azimuthal $u_\varphi$
velocity components in a meridian plane on a 95$\times$66 points
grid with $2.08$~mm spatial resolution through time series of
$400$ to $27\,000$ fields regularly sampled, at frequencies from 1
to 15 Hz, depending on the turbulence intensity and the related
need for statistics. The control parameters of the studied von
K\'{a}rm\'{a}n flow are the Reynolds number $Re= \pi (f_1+f_2)
R^2/\nu$, where $\nu$ is the fluid viscosity, which controls the
intensity of turbulence and the rotation number
$\theta=(f_1-f_2)/(f_1+f_2)$, which controls the asymmetry of the
forcing conditions. The rotation frequencies $f_{1,2}$ are
regulated by servo loop control and we obtain for $\theta$ a
typical absolute precision of $1\permil$ and time fluctuations of
the order of $\pm 2 \times 10^{-4}$. The correlation between these
fluctuations and the flow-dynamics are negligible.

\begin{figure}
\includegraphics[width=0.8\columnwidth]{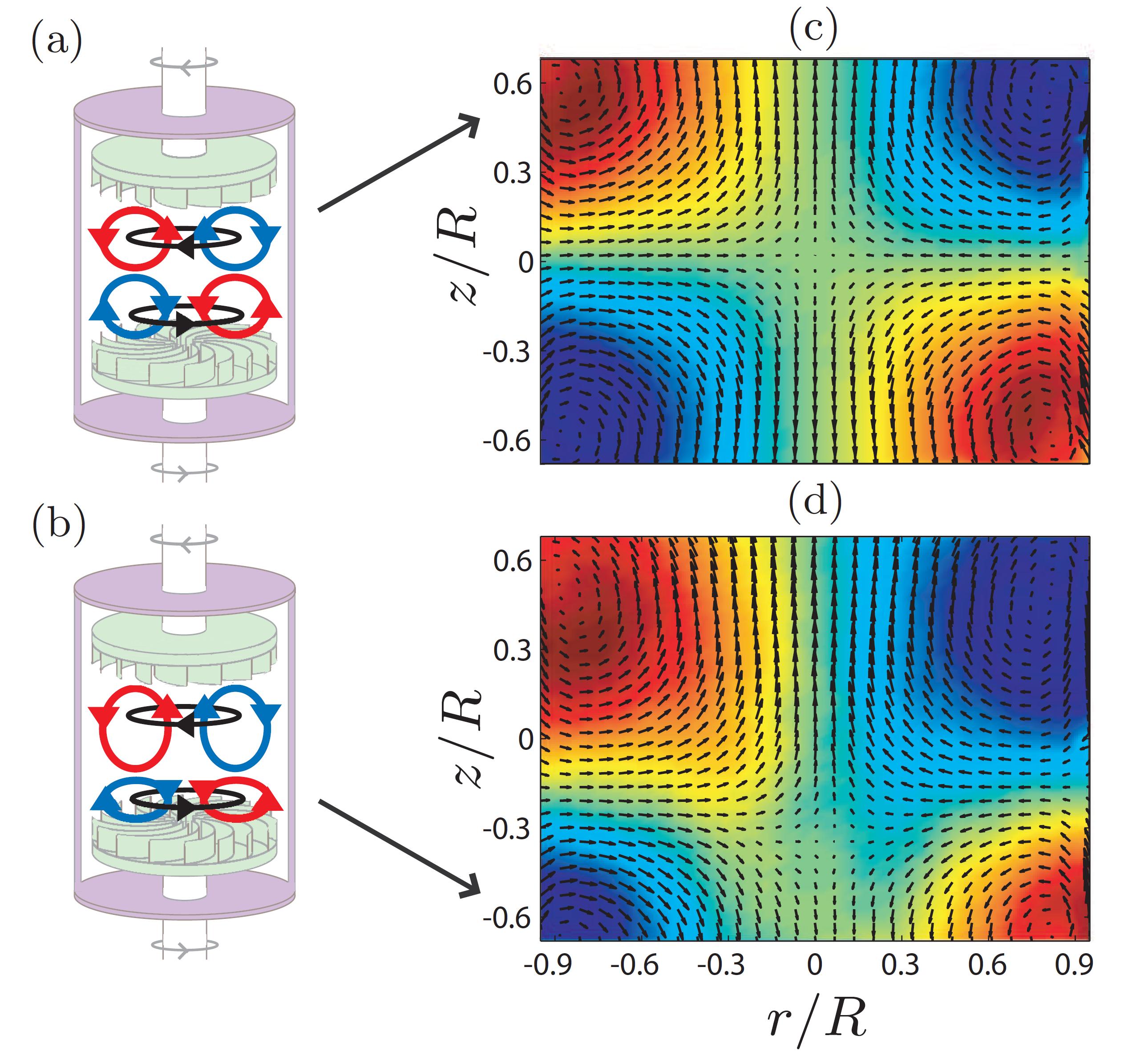}
\caption{(a) and (b) Schematic drawings of the flow topology and
(c) and (d) corresponding experimental maps of mean velocity field
of the turbulent von K\'arm\'an flow at $Re=800\,000$ for (c)
$\theta=0$ ($\overline{I}=0$) and (d) $\theta=-0.0147$
($\overline{I}<0$). The color maps the azimuthal velocity
$u_{\varphi}$, from blue to red (``jet'' colormap), whereas the
arrows map the $(u_r,u_z)$ field. The resolution has been reduced
by a factor 2 for better visibility. The $r \leftrightarrow -r$
symmetry in (c) and (d) reveals that the time-averaged mean fields
are axisymmetric. }\label{fig2}
\end{figure}

When $\theta=0$, the experimental system  is symmetric with
respect to any $\cal{R}_\pi$-rotation exchanging the two
impellers: the problem conditions are invariant under
$\pi$-rotation around any radial axis passing through the center
of the cylinder (Fig. \ref{fig1}). The symmetry group for such
experimental system is $O(2)$ \cite{nore2003}. When $\theta \ne
0$, the experimental system is no more $\cal{R}_\pi$-symmetric,
the symmetry switching to the $SO(2)$ group of rotations. However,
the parameter $\theta$, when small but non-zero, can be considered
as a measure of the distance to the exact $O(2)$ symmetry: the
\emph{stricto sensu} $SO(2)$ system at small $\theta$ can be
considered as a slightly broken $O(2)$ system
\cite{chossat93,porter2005}. Depending on the value of $\theta$,
the flow can respond by displaying different symmetries: (i) the
exact $\cal{R}_\pi$-symmetric flow composed of two toric
recirculation cells separated by an azimuthal shear layer located
at $z=0$ when $\theta=0$ (Figs. \ref{fig2}(a) and (c)); (ii) an
asymmetric two-cells flow, the shear layer being closer to the
slowest impeller ($z\ne 0$), when $\theta \ne 0$ (Figs.
\ref{fig2}(b) and (d)); (iii) and finally, a fully non symmetric
one-cell flow, the whole shear layer being concentrated in between
the blades of the slowest impeller, when $\theta$ becomes large
enough \cite{ravelet2004,ravelet2008,cortet09}.

\begin{figure}
\centerline{\includegraphics[width=\columnwidth]{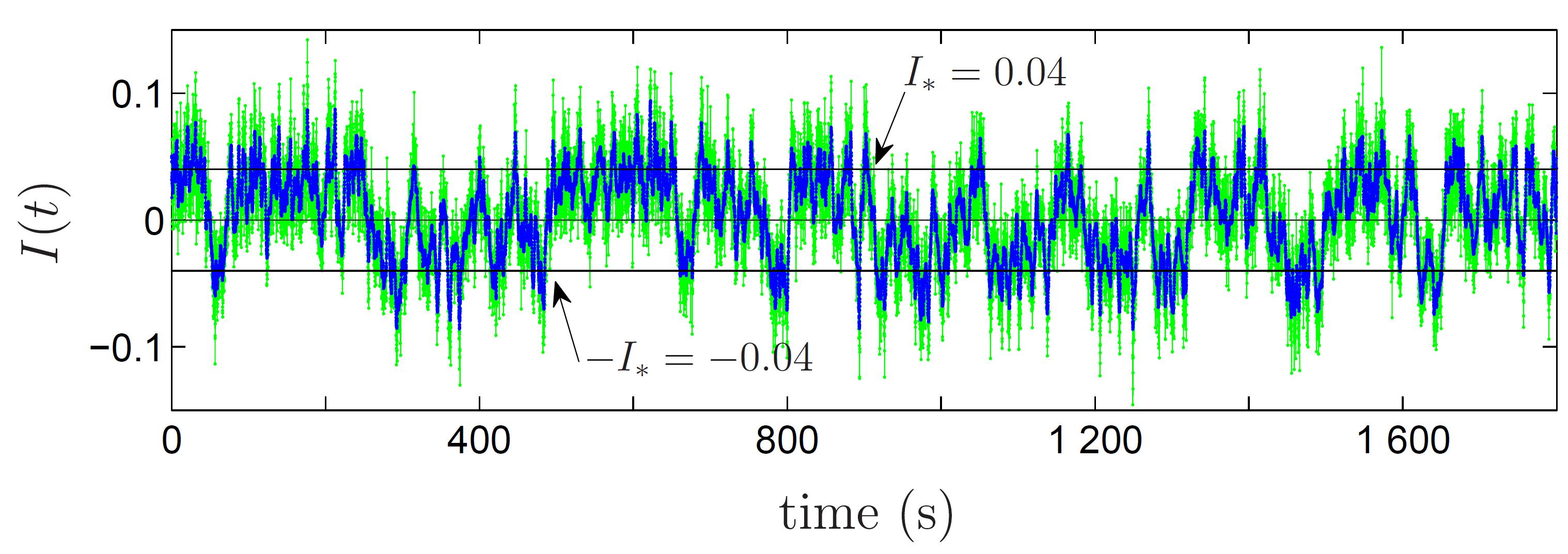}}
\caption{Global angular momentum $I(t)$ as a function of time for
an experiment performed at $Re=43\,000$ for $\theta=0$.
Green/light gray dots are PIV data sampled at 15 Hz and blue/dark
gray dots correspond to 1 Hz low-pass filtered data $I_f(t)$.
Eye-guiding lines have been drawn at $I(t)=\pm I_*=\pm 0.04$.
}\label{fig3}
\end{figure}

In order to quantify the distance of the flow to the
$\cal{R}_\pi$-symmetry, we use the normalized and space-averaged
angular momentum $I(Re,\theta,t)$ as order parameter:
\begin{equation*}
I(t)=\frac{1}{\cal{V}}\int_{\cal{V}} rdrd\varphi dz\, \frac{r u_{\varphi}(t)}{\pi \, R^2 \,(f_1+f_2)}
\label{eq:I}
\end{equation*}
where $\cal{V}$ is the volume of the flow \cite{note1}. An example
of time variation of $I(t)$ at $\theta = 0$ in the turbulent
regime is provided in Fig. \ref{fig3}. We assume that ergodicity
holds, meaning that the instantaneous turbulent flow is exploring
in time its energy landscape according to its statistical
probability. In this framework, the time average value
$\overline{I}$ of $I(t)$ is equivalent to a statistical mechanics
ensemble average providing the average is performed over a long
enough duration in order to correctly sample the slowest
time-scales. Then, using this ensemble averaged order parameter,
we define a susceptibility of the flow to symmetry breaking
$\chi_I$ as:
\begin{equation*}
\chi_I=\left.\frac{\partial \overline{I}}{\partial \theta}\right\vert_{\theta=0},
\end{equation*}
Note that $\overline{I}$ is proportional to the mean altitude
$z_s$ of the shear layer which is the natural measure of the flow
symmetry. Contrary to $z_s$, $I(t)$ is defined for any
instantaneous velocity field, including turbulent ones.

In the non-fluctuating laminar case, when $\theta=0$,
$\overline{I}=0$ due to the symmetry of the flow. In contrast, as
$\theta$ drifts away from $0$, the value of the angular momentum
$\overline{I}$ becomes more and more remote from zero as the
asymmetry of the flow grows. In such a framework, there is a
formal analogy between ferromagnetic and turbulent systems. For
ferromagnetism ({\it resp.} turbulence), the order parameter is
the magnetization $M(T,h)$ ({\it resp.} the angular momentum
$I(Re,\theta)$); the symmetry breaking parameter is the external
applied field $h$ ({\it resp.} the relative driving asymmetry
$\theta$); the control parameter is the temperature $T$ ({\it
resp.} the Reynolds number $Re$, or a function of it).

\begin{figure}
\includegraphics[width=1\columnwidth]{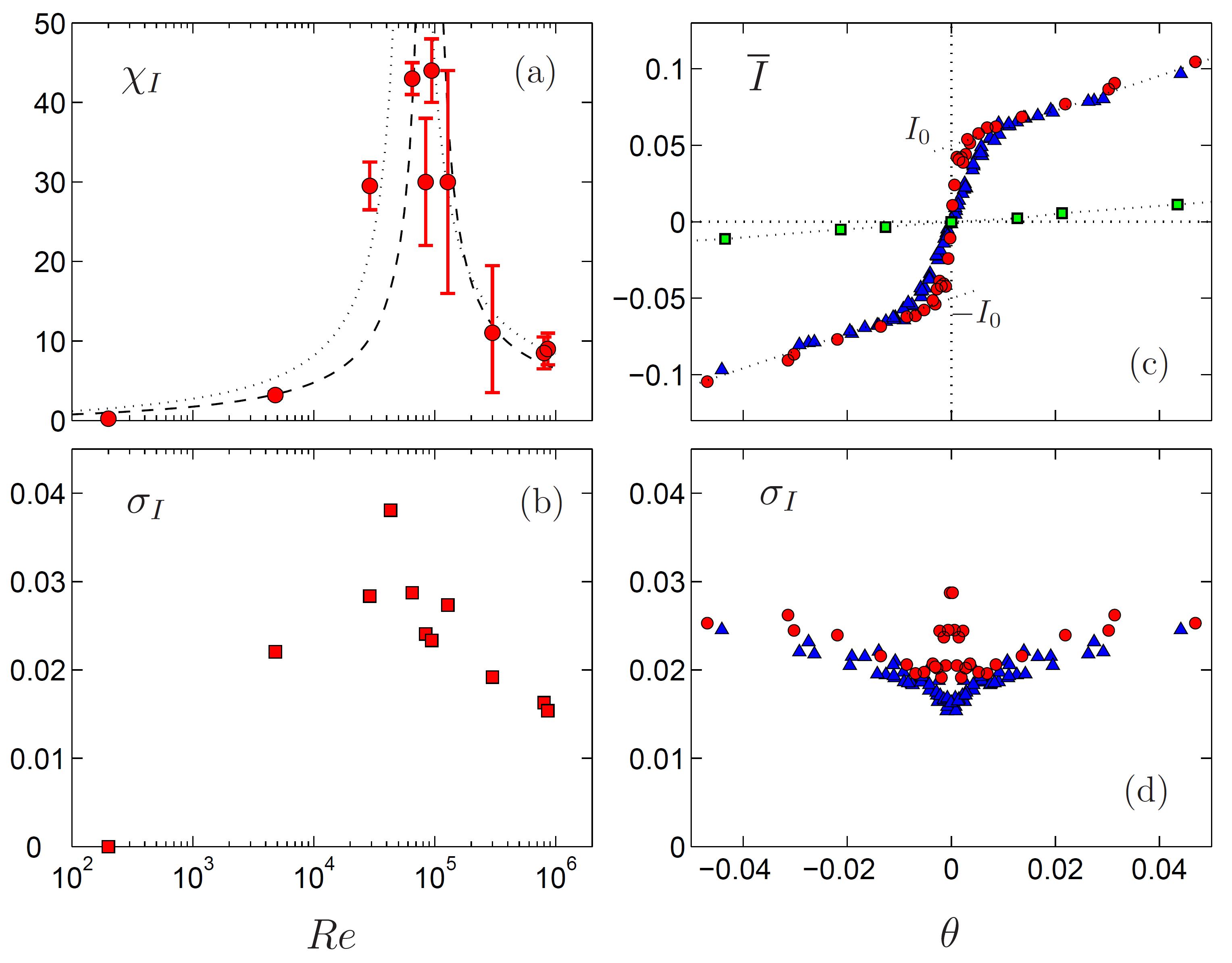}
\caption{Reynolds number dependence of: (a)  Susceptibility to
symmetry-breaking $\chi_I$ of the von K\'arm\'an mean flow at
$\theta=0$ and (b) standard deviation $\sigma_I$ of the global
angular momentum $I(t)$ at $\theta=0$. In (a), the dotted line
corresponds to the mean field theory approach with a critical
Reynolds number $Re_\chi =70\,000$ and the dashed line to $Re_\chi
= 90\,000$; (c) global angular momentum $\overline{I}$ and (d)
standard deviation $\sigma_I$ as a function of $\theta$ for
$Re=150$ (green $\Box$), $Re=65\,000$ (red $\circ$) and
$Re=800\,000$ (blue $\triangle$). In (a) and (b), horizontal error
bars are of the order of the marker size. In (a) vertical error
bars are computed as the maximum error that can arise from the
sharp dependence of $\sigma_I$ with $\theta$ around
$\theta=0$.}\label{fig4}
\end{figure}

In the sequel, we first investigate the influence of turbulence on
$\overline{I}$ and $\chi_I$ as $Re$ increases from $150$ to
$10^6$. In the laminar flow at $Re=150$, the symmetry parameter
$\overline{I}=I(t)$ evolves linearly with $\theta$ (Fig.
\ref{fig4}(c)) and the susceptibility is $\chi_I=0.240 \pm 0.005$.
Increasing the Reynolds number, one expects to reach fully
developed turbulence around $Re=10\,000$ \cite{ravelet2008}. In
such turbulent regimes, velocity fields of von K\'arm\'an flows
are characterized by a high level of intrinsic fluctuations,
\emph{i.e.} fluctuations of the same order of magnitude than the
mean values \cite{cortet09}. Therefore, even when $\theta=0$, the
$\cal{R}_\pi$-symmetry is of course broken for the instantaneous
flow. However, as usually observed for classical turbulence, this
symmetry is restored for the time-averaged flow (Fig.
\ref{fig2}(c)), which proves that time averages are long enough to
correctly sample the slowest flow time scales. Then, as in the
case of the laminar flow, when $\theta$ is varied, we observe the
breaking of the $\cal{R}_\pi$-symmetry of the mean flow (Fig.
\ref{fig2}(d)).

In Fig. \ref{fig4}(c), we see that, at $Re=800\,000$, in the close
vicinity of $\theta=0$, $\overline{I}(\theta)$ evolves actually
much more rapidly with $\theta$ than in the laminar case with the
susceptibility being larger by more than one order of magnitude:
$\chi_I=9 \pm 1$. Therefore, turbulence seems to enhance
dramatically the sensitivity of the flow to symmetry breaking.
Furthermore, for intermediate $Re=65\,000$, the slope of
$\overline{I}(\theta)$ around $\theta=0$ is even much steeper
---$\chi_I=43 \pm 1$--- than for $Re=800\,000$. In
Fig.~\ref{fig4}(a), we plot the susceptibility with respect to
$Re$. We see that the susceptibility actually grows by more than
two orders of magnitude ---from $0.24$ to $46$--- between $Re=150$
and $Re \simeq 90\,000$, before decreasing by a factor 4 between
$Re \simeq 90\,000$ and $Re=800\,000$. These results suggest a
critical behaviour for $\chi_{I}(Re)$ near $Re = Re_\chi =90\,000
\pm 10\,000$: a divergence ---revealing a continuous second order
phase transition--- or only a maximum ---revealing either a
subcritical bifurcation or a continuous transition with
finite-size effect. This cannot be experimentally tested further
since the highest measured $\chi_I$ are already of the order of
the highest measurable value considering the $\theta$ precision of
our setup. For higher $|\theta|$, we observe a crossover ---for
the slope--- in the curve $\overline{I}(\theta)$ at $|\theta_r|=(6
\pm 1)\times 10^{-3} $ for $Re=800\,000$ and very close to
$\theta=0$, at $|\theta_r|=(0.9\pm 0.15)\times 10^{-3}$, for
$Re=65\,000$ (Fig. \ref{fig4}(c)). For $|\theta|>|\theta_r|$, we
recover the laminar slow evolution of $\overline{I}$ with $\theta$
up to $\theta=\pm 0.1$ where the flow bifurcates to the one-cell
topology (not shown). Since $\overline{I}(\theta)$ is quite
independent of $Re$ for $|\theta|>|\theta_r|$ at large $Re$, we
can extrapolate this linear behaviour to $\theta=0$. This
extrapolation describes the ideal behaviour at critical Reynolds
number $Re_\chi$ if $\chi_I$ diverges:  a jump of $\overline{I} $
between $-I_0$ and $+I_0$ where $I_0 \simeq 0.05$. This can be
interpreted as a spontaneous ``turbulent momentization'' $I_0$ at
$\theta=0$
---possibly affected by finite-size effects--- by analogy with the
spontaneous magnetization $M_0$ at zero external field for
ferromagnetism. It is also similar to the experimental results of
\cite{Torre2007}.

A signature of this momentization can be seen on the instantaneous
global angular momentum $I(t)$ for $Re$ near the peak of
susceptibility and $\theta=0$ (\textit{e.g.} in Fig. \ref{fig3}).
Indeed, one observes that $I(t)$ does not remain near zero (its
mean value) but shows a tendency to lock preferentially on the
plateaus $\pm I_*$ with $I_* = 0.04 \simeq I_0$. Therefore, the
turbulent flow explores a continuum of metastable symmetry
breaking patterns evidenced by $-0.04 \lesssim I_f(t) \lesssim
0.04$, $I_f(t)$ being the 1 Hz low-pass filtered value of $I(t)$.
The global angular momentum actually fluctuates very much along
time with two separate time scales: fast fluctuations related to
``traditional'' small scale turbulence and time intermittencies
corresponding to residence time of few tens of seconds. If one
performs a time average over one of these intermittent periods
only, one obtains a time localized ``mean'' flow, which breaks
spontaneously the symmetry, analogous to what is obtained for true
mean flows when $\theta\neq0$ as presented in Figs. \ref{fig2}(b)
and (d).

The presence of strong fluctuations is not surprising here: close
to a phase transition we expect critical fluctuations. To check
this, we compute the standard deviation $\sigma_I(Re,\theta)$.
Fig.~\ref{fig4}(b) shows, for $\theta=0$, how $\sigma_I$ varies
from zero in laminar case to finite values for highly turbulent
flows going through a maximum at Reynolds number $Re =
Re_\sigma=45\,000 \pm 10\,000$ located below the peak of
susceptibility at $Re_\chi$. Additionally, in Fig. \ref{fig4}(d),
the dependence of $\sigma_I$ as a function of the symmetry control
parameter $\theta$ reveals a strong difference between the two
Reynolds numbers shown: $\sigma_I(\theta)$ presents a sharp and
narrow peak at $\theta=0$ for $Re=65\,000$, which does not exist
for $Re=800\,000$. The amplitude of this peak from its bottom to
its top actually measures the additional amount of low frequency
symmetry fluctuations due to the multistability. This amount of
fluctuations appears to be connected to the susceptibility
increase below $Re_\chi$.

The previous experimental results set a strong connection between
the spontaneous symmetry fluctuations of the flow and the mean
flow response to the system symmetry breaking: the interpretation
of the large fluctuations of $I(t)$ in terms of multistability
suggests that the strong observed linear response of the mean flow
(Fig. \ref{fig4}(c)) with respect to $\theta$ in the close
vicinity of $\theta=0$ is the result of a temporal mixing between
the metastable states in different proportions.

Nevertheless, as the Reynolds number is varied, two distinct
maxima have been evidenced in our turbulent flow: one for the
susceptibility $\chi_I$ to the $O(2)$-symmetry breaking near
$Re=Re_\chi \simeq 90\,000$, and the other for the standard
deviation of the global angular momentum $I(t)$ near $Re=
Re_\sigma \simeq 45\,000$. In this Reynolds number range, the
turbulence is generally expected to be fully developed,
\emph{i.e.} any non-dimensional characteristic quantity of the
flow should be $Re$-independent. This is definitively not the case
in this von K\'arm\'an experiment. Actually, visual observations
of the flow reveal an increase of the average azimuthal number $m$
of large scale vortices in the shear-layer from $m=3$ to $m=4$
through an Eckhaus-type transition, between $Re_\sigma$ and
$Re_\chi$. In the following, we make the hypothesis that at least
one critical phenomenon exists in the range $50\,000 \apprle Re
\apprle 100\,000$. Using the turbulent von
K\'arm\'an-ferromagnetism analogy, we can check how classical mean
field predictions for second order phase transition apply to our
system. In terms of susceptibilities, it predicts a critical
divergence at a temperature $T_c$: $\chi \propto \left \vert T-T_c
\right \vert^{-1}$. Since the logarithm of the Reynolds number has
already been proposed as the control parameter governing the
statistical temperature of turbulent flows ---$T\sim 1/\log Re$
\cite{castaingbis}--- , this prediction translates in our case
into:
\begin{equation*}
\chi_I \propto \left \vert 1/\log Re- 1/\log Re_c\right \vert^{-1}.
\label{MFanalog}
\end{equation*}
This formula reasonably describes our data with $Re_c$  between
typically $70\,000$ and $90\,000$ (Fig.  \ref{fig4}(a)) supporting
the asymmetry between the two branches even with a unique exponent
$-1$. As far as fluctuations are concerned, it is difficult to
find a reasonable critical exponent for $\sigma_I$. However, the
results show that the maxima for $\chi_I$ and $\sigma_I$ are
clearly separated. This is at variance with classical phase
transition theory, stating \emph{e.g.} in the Fluctuation
Dissipation Theorem that $\sigma_I^2$ should be proportional to
$\chi_I$. A reason for this discrepancy lies in the high level of
intrinsic fluctuations in our system that makes this transition
non-classical. Instabilities or bifurcations occurring on highly
fluctuating systems are commonly found in natural systems and the
literature reports transitions and symmetry breaking at high
Reynolds number (see, \emph{e.g.}, references
\cite{mujica,gibert2009,castaing,lohse2009,Torre2007,tabeling96,tabeling02,dynamoP1,dynamoP5,ravelet2004,ravelet2008})
but the corresponding theoretical tools are still today not well
settled. Existing studies of phase transitions in the presence of
fluctuations generally considers systems in which an external
noise ---additive or multiplicative---  is introduced
\cite{review_noise}. In particular, it has been shown in models
that multiplicative noise can produce an ordered symmetry breaking
state through a non equilibrium phase transition
\cite{vandenbroeck94}. This behaviour could be at the origin of
our observed transition. Finally, we can notice that, as the
Reynolds number increases, the \lq\lq turbulent
momentization\rq\rq $I_0$ first increases and then decreases,
contrary to the magnetization in the usual para-ferromagnetic
transition. This result is reminiscent of a reentrant
noise-induced phase transition similar to that observed in the
annealed Ising model \cite{thorpe76,genovese98}. The study of the
evolution of $I_0$ and/or $I_*$ with $Re$ requires more statistics
and is left for future work. Finally, our turbulent system, in
which we can have access both to the spatiotemporal evolution of
the states and to the mean thermodynamic variables, appears as a a
unique tool to study out-of-equilibrium phase transitions in
strongly fluctuating systems.

\begin{acknowledgments}
We thank K. Mallick for fruitful discussions. PPC was supported by
Triangle de la Physique.
\end{acknowledgments}


\begin{thebibliography}{10}

\bibitem{landau} L. D. Landau and E. M. Lifchitz, \textit{Statisticheskaya Fisika} (Nauka, Moscow, 1976).

\bibitem{manneville} P. Manneville, \textit{Dissipative Structures and Weak Turbulence} (Academic Press, Boston, 1990).

\bibitem{frisch} U. Frisch, \textit{Turbulence - The Legacy of A N Kolmogorov} (Cambridge University Press, Cambridge, 1995).

\bibitem{hof} B. Hof, J. Westerweel, T. Schneider and B. Eckhardt, Nature \textbf{443}, 59 (2006).

\bibitem{tabeling96} P. Tabeling \textit{et al.}, Phys. Rev. E \textbf{53}, 1613 (1996).

\bibitem{tabeling02} P. Tabeling and H. Willaime, Phys. Rev. E \textbf{65}, 066301 (2002).

\bibitem{ravelet2004} F. Ravelet, L. Mari\'e, A. Chiffaudel and F. Daviaud, Phys. Rev. Lett. \textbf{93}, 164501 (2004).

\bibitem{castaing} F. Chill\'{a}, M. Rastello, S. Chaumat and B. Castaing, Eur. Phys. J. B \textbf{40}, 223 (2004).

\bibitem{mujica} N. Mujica and D. P. Lathrop, J. Fluid Mech. \textbf{551}, 49 (2006).

\bibitem{ravelet2008} F. Ravelet, A. Chiffaudel and F. Daviaud, J. Fluid. Mech. \textbf{601}, 339 (2008).

\bibitem{lohse2009} R. Stevens \textit{et al.}, Phys. Rev. Lett. \textbf{103}, 024503 (2009).

\bibitem{nore2003} C. Nore, L. S. Tuckerman, O. Daube and S. Xin, J. Fluid Mech. \textbf{477}, 1 (2003).

\bibitem{chossat93} P. Chossat, Nonlinearity \textbf{6}, 723 (1993).

\bibitem{porter2005} J. Porter and E. Knobloch, Physica D \textbf{201}, 318 (2005).

\bibitem{cortet09} P.-P. Cortet \textit{et al.}, Phys. Fluids \textbf{21}, 025104 (2009).

\bibitem{Torre2007} A. de la Torre and J. Burguete, Phys. Rev. Lett. \textbf{99}, 054101
(2007); J. Burguete and A. de la Torre, Int. J. Bif. and Chaos
\textbf{19}, 2695 (2009).

\bibitem{castaingbis} B. Castaing, J. Phys. II \textbf{6}, 105 (1996).

\bibitem{gibert2009} M. Gibert \textit{et al.}, Phys. Fluids \textbf{21}, 035109 (2009).

\bibitem{dynamoP1} R. Monchaux \textit{et al.}, Phys. Rev. Lett. {\bf 98},  044502 (2007).

\bibitem{dynamoP5} R. Monchaux \textit{et al.}, Phys. Fluids {\bf 21}, 035108 (2009).

\bibitem{review_noise} N. G. van Kampen, \textit{Stochastic Processes in Physics and Chemistry} (North-Holland
Personal Library, Elsevier, Amsterdam, 1981).

\bibitem{vandenbroeck94} C. Van den Broeck, J. Parrondo and R. Toral, Phys. Rev. Lett. {\bf 73}, 3395 (1994).

\bibitem{thorpe76} M. Thorpe and D. Beeman, Phys. Rev. B {\bf 14}, 188 (1976).

\bibitem{genovese98} W. Genovese, M. Munoz and P. Garrido, Phys. Rev. E {\bf 58}, 6828 (1998).

\bibitem{note1} Practically, $I(t)$ is computed from PIV-data restricted
to a meridian plane only but, since azimuthal flow fluctuations
are strong, time-average over several rotation periods
---statistically equivalent to spatial azimuthal averaging---
estimate correctly the 3D-value of $I(t)$ (see details in
Ref.~\cite{cortet09}).

\end{thebibliography}
\end{document}